\documentclass[%
 aip,
 amsmath,amssymb,
 reprint,%
]{revtex4-1}
\usepackage[T1]{fontenc}
\usepackage{newtxtext,newtxmath}
\usepackage{graphicx}
\usepackage{dcolumn}
\usepackage{bm}
\usepackage{siunitx}
\usepackage{etoolbox}
\usepackage{xcolor}
\usepackage{microtype}
\usepackage[colorlinks=true,allcolors=blue!50!black]{hyperref} 

\begin{document}

\title{Optical Speckle Patterns as Nonlinear Feature Maps for Photonic Machine Learning}

\author{Kiia Arola}
\email{kiia.arola@tuni.fi}
\affiliation{Photonics Laboratory, Physics Unit, Tampere University, Tampere, Finland}

\author{Mikko J. Huttunen}
\affiliation{Photonics Laboratory, Physics Unit, Tampere University, Tampere, Finland}

\date{\today}

\begin{abstract}
The rapid growth of artificial intelligence (AI) has increased the demand for en\-er\-gy-ef\-fi\-cient hardware, and optical computing has emerged as a promising alternative to conventional electronic processors owing to its high parallelism and potentially lower power consumption. In this work, we investigate an optical computing architecture based on an integrated photonic circular multipass cell, focusing on its suitability for ma\-chine learning and pattern recognition tasks. The properties of the system are evaluated through measures of nonlinearity and cosine similarity, which characterize the system's information-processing capacity and the dimensionality of the generated feature representations. Numerical results demonstrate that the reservoir mirror reflectance, the size of the encoded figure, and the refractive-index contrast between the reservoir and the encoded figure modify the nonlinearity of the system, as verified via pairwise similarity estimates. These results indicate that the features generated by the circular multipass cell can be tailored by adjusting the system parameters, providing a means to control the reservoir transformation of the input data. The resulting representations were applied to classification tasks involving tex\-ture images and thyroid collagen images, achieving classification accuracies ranging from 69\% to 93\%. These findings suggest that photonic computing systems could provide a basis for more en\-er\-gy-ef\-fi\-cient ma\-chine-learning hardware.
\end{abstract}


\maketitle

\section{Introduction}

Many classification problems involve datasets that are not linearly separable in their original feature space \cite{CortesCorinna1995Sn}. In such cases, performance can be improved by transforming the data into a higher-dimensional representation in which simple linear decision boundaries become effective. This principle underlines a wide range of ma\-chine-learning methods, including kernel approaches \cite{Scholkopf2002}, neural networks \cite{LeCun2015}, and reservoir computing (RC) \cite{VandoorneKristof2008Tosp,VanderSande2017,Tanaka2019,Appeltant2011}.

Optical computing provides an attractive platform for performing such transformations directly in the physical domain \cite{DenisZakari2022PKML,VanderSande2017,HuJingtian2024Docf,Yildirim2024Npl}. Rather than computing nonlinear mappings digitally, optical systems exploit light propagation and light--matter interactions to generate complex feature transformations, potentially with ultrafast speeds and low energy consumption \cite{Rafayelyan2020LORC,Aifer2025,HuJingtian2024Docf,Yildirim2024Npl}. Various implementations have been demonstrated, including optical nonlinearities~\cite{Yildirim2023,TeginUgur2021Solo,Stiller2024Pcsw}, free-space optical systems \cite{Rafayelyan2020LORC,RefJ1,HanSeungYun2025Okmw,McMahon2024,HuJingtian2024Docf,LiJingxi2022Pmc}, and integrated photonic architectures \cite{wang2024ultrafast,WanjuraClara2024Fnnc,Cem2023Dmm}.

Among these approaches, recent works based on free-space optical systems have shown that multiple-scat\-ter\-ing events can project input data into high-dimensional vector spaces suitable for ma\-chine-learning tasks \cite{RefJ1,HanSeungYun2025Okmw,McMahon2024,HuJingtian2024Docf,LiJingxi2022Pmc}. Such systems employ multiple-scat\-ter\-ing events to generate complex optical transformations entirely without active --- or nonlinear --- optical elements, enabling applications such as image classification \cite{McMahon2024}, time-series prediction \cite{WangHao2025Ongr,YouMengyu2025Ntsc}, image reconstruction, keypoint detection, and object detection \cite{Rafayelyan2020LORC}. Related approaches have also demonstrated that passive diffractive optical systems can implement arbitrary linear transformations through free-space propagation \cite{LiJingxi2022Pmc}, while recent work has shown that nonlinear neuromorphic computations can emerge from linear wave scat\-ter\-ing without relying on physical nonlinearities \cite{WanjuraClara2024Fnnc}.

Optical reservoir computing has previously been demonstrated using a variety of photonic platforms, including scattering-based architectures \cite{RefJ1,HanSeungYun2025Okmw,McMahon2024}. Similar speckle-based feature generation has also been proposed on photonic integrated circuits, where coherent mixing of delayed optical inputs in a passive waveguide array, followed by square-law detection, has been shown to yield nonlinear feature vectors for signal equalization \cite{Cox2026NLPO}. The same principle motivates the use of circular multipass cells (CMPCs), which increase the effective optical path length through repeated reflections within a compact geometry, thereby enabling complex optical interactions within a small footprint \cite{Graf2018}. This makes CMPCs a promising platform for implementing optical reservoir computing.

Reservoir computing is a ma\-chine-learning framework in which only a simple readout layer is trained, while the reservoir itself remains fixed~\cite{Jaeger2004,Maass2002,Lukosevicius2009}. The reservoir projects input data into a high-dimensional feature space through its internal dynamics, after which task-relevant information is extracted using supervised learning. Optical implementations of RC are particularly attractive, since the feature transformation can be performed directly by the physics of light propagation, potentially reducing computational costs compared with conventional digital approaches.

In this work, we propose an integrated architecture that implements a CMPC-like operating mechanism. For this reason, the proposed system is referred to as an integrated CMPC architecture. We numerically investigate its use as an optical reservoir for image classification. In the proposed system, repeated scattering, interference, and propagation events transform the encoded input patterns into high-dimensional optical representations. The resulting speckle patterns, assumed to be imaged by a camera positioned above the cell, serve as the reservoir states. The simulations suggest that the optically generated features can improve class separability, resulting in higher classification accuracies than what are achieved using simple linear classifiers. The optically enhanced feature transformations yielded classification accuracies of up to 93~\% for texture classification and 69~\% for the biomedical image dataset. As a substantial part of the feature extraction is performed directly in the optical domain, the proposed approach combines the principles of RC with the compactness of integrated photonics, offering a potentially scalable platform for optical information processing.

\section{CMPC Reservoir Model}

We propose and study an optical reservoir system that could be realized using various integrated photonic material platforms. In short, the system consists of a slab waveguide layer, inside which light can freely propagate as slab-waveguide modes. The system is assumed to be sufficiently large such that the propagation of light within the slab layer can be modelled using geometrical light rays. Here, we considered a CMPC structure with a diameter of 1 cm. The surrounding mirrors forming the reservoir are assumed to be realized by using curved distributed Bragg reflectors (DBRs) in the structure. Such devices could be fabricated using standard lithographic techniques.

\begin{figure*}[!t]
\includegraphics[width=\textwidth]{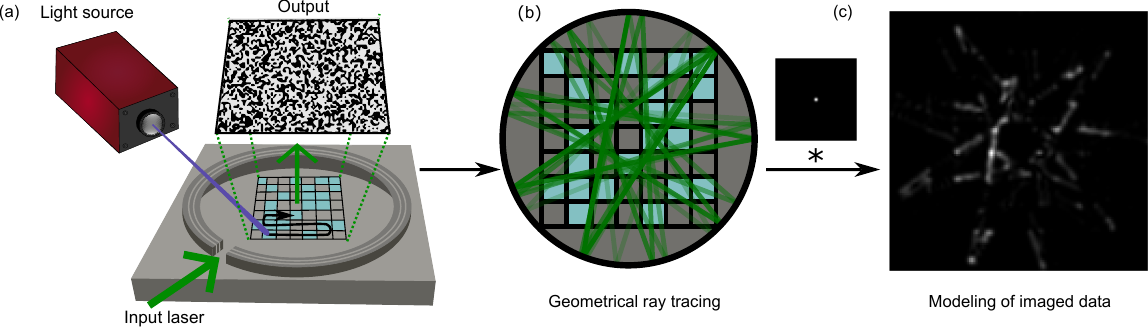}
\centering
\caption{\label{fig:NGRC_schematic}  (a) Schematic illustration of the integrated CMPC structure. (b) Geometrical ray trajectories are rasterized to generate an electric field matrix. (c) The image data are then modeled by convolving the electric field matrix with a small Gaussian kernel. The output intensity field is obtained by taking the squared magnitude of the resulting field.}
\end{figure*}

A schematic illustration of the proposed system is shown in Fig.~\ref{fig:NGRC_schematic}(a). The operating principle is based on encoding the input data as spatially varying refractive-index perturbations $\Delta n(x,y)$ within the reservoir that modify the propagation of light rays inside the slab waveguide layer. Consequently, different input patterns produce different distributions of optical trajectories within the reservoir (See Fig.~\ref{fig:NGRC_schematic}(b)). As the propagating light repeatedly interacts with the encoded refractive-index pattern $\Delta n(x,y)$ and reflects from the CMPC boundaries, the resulting optical response becomes nonlinearly dependent on the input pattern.

The refractive-index profile $\Delta n(x,y)$ can be encoded into the CMPC structure by several methods. Previous studies have demonstrated local and persistent modification of the refractive index in optical waveguides using photoactive polymer films, in which optical illumination induces spatially patterned index changes \cite{OnoderaTatsuhiro2026Acom,Yanagimoto2025}. Alternatively, the thermo-optic effect of the slab material can be exploited, either by means of integrated microheaters or via laser scanning using a wavelength that is absorbed by the slab material~\cite{Faraon2009,Harris2014}. Other approaches include free-carrier injection, e.g., by optically pumping electrons into the conduction band of the semiconductor \cite{Soref1987,Leonard2002,Fushman2007} and non-volatile phase-change materials~\cite{Wuttig2017,Delaney2021}. Together, these techniques provide feasible routes to realize refractive-index perturbations considered in this work ($\Delta n \approx 10^{-3}-10^{-1}$). We emphasize, however, that the analysis presented here is not restricted to any particular implementation.

To enable observation of the reservoir state, we further assume that the roughness of the surface of the top of the reservoir introduces weak out-of-plane scat\-ter\-ing of light from the waveguide layer. As a result, the internal optical field can be observed as a speckle pattern using an imaging camera (see Fig.~\ref{fig:NGRC_schematic}(a)). 

The simulations were performed using MATLAB and custom-developed ray-tracing algorithms. Within the CMPC, light rays were assumed to propagate within the slab waveguide in similar manner as free-space beams would propagate withing bulk CMPC structures, being reflected by the mirror walls of the CMPC~\cite{TienP.K.1977Ioan}. The encoded image was represented as a spatial refractive index distribution $\Delta n(x,y)$ located at the center of the reservoir and discretized into pixels, each corresponding either to the background refractive index of the reservoir medium or to a locally modified refractive-index value. Reservoir refractive index was chosen to be 2.011, being close to the refractive index of silicon nitride, deposited via plasma-enhanced chemical vapor deposition (PECV)~\cite{Koskinen2017,Ning2012}. Consequently, propagating rays experience varying amounts of refraction depending on the local refractive-index distribution, the reflectance of the reservoir mirrors, and the size of the encoded image, as illustrated in Fig.~\ref{fig:NGRC_schematic}(b).

During ray propagation, each time a ray encounters a re\-frac\-tive in\-dex interface was treated as a separate refraction event. The propagation direction after the interface was calculated according to Snell's law, while the corresponding amplitude change was determined using the Fresnel coefficients~\cite{TienP.K.1977Ioan}. At each interface, a portion of the optical field is reflected and a portion is transmitted. To reduce computational complexity, only the ray carrying the larger optical amplitude was retained and propagated further, whereas the weaker ray was neglected. When rays were reflected from the mirrors of the CMPC, the associated amplitude reduction due to the mirror reflectance was taken into account, and the new propagation direction was determined according to the law of reflection. In each simulation, 100 light rays were launched with an initial amplitude of 1. The propagation of an individual ray was terminated when its amplitude decreased below 0.1\% of its initial value, corresponding to an amplitude threshold of 0.001.

The output intensity distribution $I_\mathrm{out}$, acting as the data vector fed into conventional electronic classification layers, was reconstructed as follows. First, we simulated ray trajectories within the CMPC structure, discretizing the cell into a 120$\times$120 matrix. In the ray-tracing model, each ray trajectory consists of straight-line segments. To convert these trajectories into a matrix representation, each segment was rasterized using the Bresenham algorithm~\cite{Wetzler2010Bresenham}. The complex electric field in each grid cell was then obtained by coherently summing the contributions of all rays passing through that cell, \begin{equation}
E(x,y)=\sum_j E_{0,j}e^{i\phi_j}, \qquad \phi_j=kL_j,
\end{equation}
where $E_{0,j}$ is the amplitude of the $j$th ray, $k$ is the wavenumber, and $L_j$ is the optical path length accumulated by the ray from its source to the corresponding cell. Possible phase shifts due to mirror reflections were neglected.


To model the experimentally observed speckle pattern, each point along the ray trajectories was treated as a point source emitting a spherical wave, thereby accounting for the spreading of the rays during free-space propagation toward the camera. This corresponds to a convolution of the source field with a translation-invariant propagation kernel~\cite{novotny2012}, approximated by a Gaussian function with width $\sigma$ representing the scattered field on the camera
\begin{equation} I_\mathrm{out}(X,Y)=\left|E(x,y)*K(x-X,y-Y)\right|^2, \end{equation} where $K(x-X,y-Y)= \exp\!\left[ -\frac{(x-X)^2+(y-Y)^2}{2\sigma^2} \right]$ and $*$ denotes convolution. The resulting speckle pattern, which in an experiment would be recorded with a camera, is shown in Fig.~\ref{fig:NGRC_schematic}(c). This speckle pattern was used in all subsequent analyses, including the nonlinearity, cosine-similarity, and classification studies.

\section{The Nonlinear Capabilities}\label{nonlinearity}

A key requirement for optical computing is a nonlinear relationship between the input and output signals. To investigate the nonlinear response of the CMPC, the input refractive-index distribution $\Delta n(x,y)$ was represented by a binary 7\(\times\)7 grid. Each pixel corresponded either to the refractive index of the reservoir or to a predefined refractive-index perturbation, resulting in \(2^{49}\) possible input configurations. Due to the very large number of possible configurations, only a smaller subset was used in simulations. Here, numerical simulations are shown for randomly selected 130 configurations for numerical simulations.

Encouraged by recent work \cite{HanSeungYun2025Okmw}, the degree of nonlinearity was quantified by comparing the simulated output intensity fields with a fitted linear model. For this purpose, the output intensity vectors corresponding to all simulated input patterns were collected into a single matrix,
\begin{equation}\label{all_Is}
    I_\mathrm{out}=[I_{\mathrm{out},1},I_{\mathrm{out},2},\ldots,I_{\mathrm{out},N}],
\end{equation}
where each column represents the output intensity vectors associated with a specific input pattern. A linear model, \(\hat{I}_{\mathrm{out}} = Wx\), was then fitted, where \(x\) contains the binary input patterns and \(W\) is the fitted weight matrix. The binary input vectors were constructed such that pixels with a refractive index different from that of the reservoir were assigned a value of 1, while pixels sharing the reservoir refractive index were assigned a value of 0. The agreement between simulated \(I_\mathrm{out}\) and fitted outputs was quantified using the coefficient of determination~\cite{HanSeungYun2025Okmw}
\begin{equation}
R^2 = 1 - \frac{\mathrm{Var}\left(I_{\mathrm{out}}-\hat{I}_{\mathrm{out}}\right)}
{\mathrm{Var}\left(I_{\mathrm{out}}\right)}.
\end{equation}
Lower \(R^2\) values indicate stronger nonlinear behavior, whereas values approaching unity correspond to a linear dependence.

\begin{figure*}[!t]
\includegraphics[width=\textwidth]{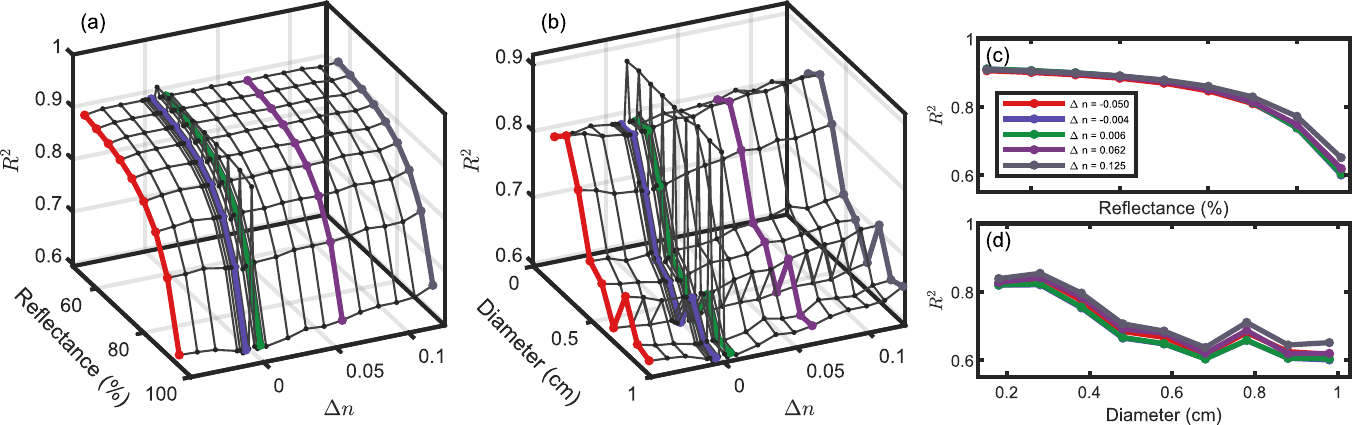}
\centering
\caption{\label{fig:R2} Coefficient of determination (\(R^2\)) as a function of refractive-index contrast and (a) Bragg-mirror reflectance or (b) encoded-image diameter. Selected refractive-index contrast values are highlighted by color and replotted in (c) and (d), showing the dependence of \(R^2\) on reflectivity and encoded-image diameter, respectively.}
\end{figure*}

Figure~\ref{fig:R2}(a)--(b) illustrates the dependence of the \(R^2\) values on mirror reflectivity $ \mathcal{R}$, encoded-image diameter $d$, and refractive-index contrast $\Delta n$. A set of evenly spaced refractive-index contrast values was selected from each dataset and highlighted using different colors. The corresponding \(R^2\) curves for the selected contrasts are shown separately in Figs.~\ref{fig:R2}(c)--(d).

The effect of mirror reflectivity $\mathcal{R}$ is shown in Fig.~\ref{fig:R2}(a). Increasing the reflectivity generally decreases the coefficient of determination \(R^2\), indicating stronger nonlinear transformations within the CMPC reservoir. This behavior can be attributed to the effectively increased interaction time of light (or optical path length).

A similar trend can be observed for the encoded-image diameter $d$ in Fig.~\ref{fig:R2}(b). As the image diameter increases, the \(R^2\) value decreases, indicating an increasing departure from the linear response. Larger encoded images occupy a greater fraction of the reservoir and increase the effective propagation distance, allowing phase perturbations to accumulate over longer optical paths. Consequently, the generated output intensity fields become increasingly complex.

Compared with the trends associated with reflectivity and image diameter, the influence of the refractive-index contrast on the $R^2$ values appears relatively weak. This nonlinear behavior can be attributed to the short wavelength, as becomes clear upon repeating the simulations at very long wavelengths (see supplementary material, section~\ref{chap:effects_wavelength} for details). Although the smallest refractive index contrast considered here is \(10^{-3}\), additional simulations (data not shown) indicate that a contrast as small as \(10^{-4}\) is already sufficient to make the system's \(R^2\) value deviate strongly from unity. This behavior can be seen more clearly in Figs.~\ref{fig:R2}(c)--(d), where the \(R^2\) values corresponding to different refractive-index contrasts are shown. Despite the range of contrast values considered, the resulting curves remain closely grouped together.

\section{CMPC Feature-Space Analysis}

In the previous section we observed that the system performs a feature transformation, which induces a nonlinear relationship between the input patterns and the output intensity fields.  However, changes in $R^2$ alone do not reveal whether the response corresponds to a genuinely rich\-er feature space or simply produces output patterns that appear different while still carrying largely redundant information~\cite{Legenstein2007EdgeOfChaos}. To address this question, we examined the similarity of output intensity vectors using pairwise cosine similarity under same system parameters. The cosine similarity between two output vectors ${I}_i$ and $I_j$ was calculated as~\cite{Salton1975VSM}
\begin{equation}
\cos(I_i,I_j) = \frac{I_i^{T}I_j} {\|I_i\|\,\|I_j\|}, \label{eq:cosine_similarity}
\end{equation}
where values close to unity indicate highly similar responses and smaller values correspond to increasingly distinct output representations. Cosine similarity is wide\-ly used to quantify similarity between high-dimensional vectors because it is insensitive to overall signal magnitude and depends only on the relative orientation of the vectors in feature space \cite{Salton1975VSM,Manning2008IR,Kulce2021}.


\begin{figure}[!t]
    \centering
    \includegraphics[width=\columnwidth]{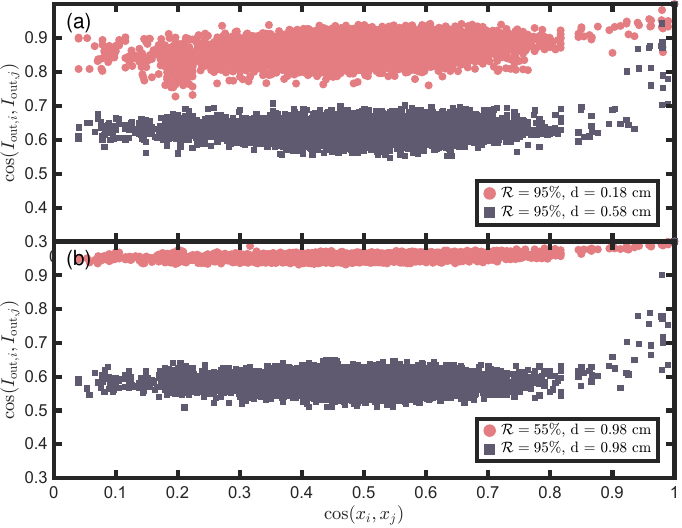}
    \caption{Pairwise cosine similarity between the binary input patterns and the corresponding output intensity fields as a function of (a) the encoded-image diameter at a fixed reservoir-mirror reflectance of 95\%, and (b) the reservoir-mirror reflectance at a fixed encoded-image diameter of 0.98 cm. The refractive-index contrast is 0.01 in both cases.}
    \label{fig:sim}
\end{figure}
Figure~\ref{fig:sim}(a) shows that varying the diameter of the encoded feature changes the distribution of output cosine similarities. With the refractive-index contrast fixed at 0.01 and the mirror reflectivity fixed at 95\%, a diameter of $d=0.18$~cm yields output pairs with cosine similarities concentrated between approximately 0.80 and 0.90, whereas increasing the diameter to $d=0.58$~cm shifts the distribution to lower values, around 0.55--0.65. Figure~\ref{fig:sim}(b) shows an analogous effect for the mirror reflectivity, with the refractive-index contrast again fixed at 0.01 and the encoded feature diameter fixed at $d = 0.98$~cm. A reflectivity of $\mathcal{R}=55$~\% produces output similarities close to unity, while increasing the reflectivity to $\mathcal{R}=95$~\% substantially lowers the similarity values, to around 0.6.
%

A potential explanation for these observations is that, when the refractive index contrast is small, each scattering or refraction event introduces only a minor perturbation to the ray trajectory. As a result, rays generated by different input patterns follow similar propagation paths during the early stages of propagation, and differences emerge only after multiple interactions as these small deviations accumulate. In addition, for smaller encoded input patterns, rays are less likely to intersect the structure, reducing its influence on the output intensity vectors and increasing the similarity between the generated output patterns.

Overall, the results indicate that the richness of the feature space generated by the CMPC can be controlled through the encoded figure size $d$ and mirror reflectivity $\mathcal{R}$, both of which affect the effective number of scat\-ter\-ing and cavity-interaction events experienced by the propagating field. However, the pairwise cosine similarity approaches a nonzero asymptotic value as these parameters are increased, suggesting that the generated feature space still contains a degree of redundancy. In other words, additional scat\-ter\-ing and cavity interactions do not continuously produce entirely independent features but instead yield partially correlated output patterns. We note, that as seen in above results and simulations shown in the supplementary material, the system becomes nonlinear (\(R^2\) deviates from unity) already with very small refractive index perturbations. However, greater changes in the index contrast didn't anymore markedly affect the nonlinearity, or clearly affect the system's capabilities for computing.

\section{Classification demonstrations}
 
\begin{figure*}[!t]
    \centering
    \includegraphics[width=\textwidth]{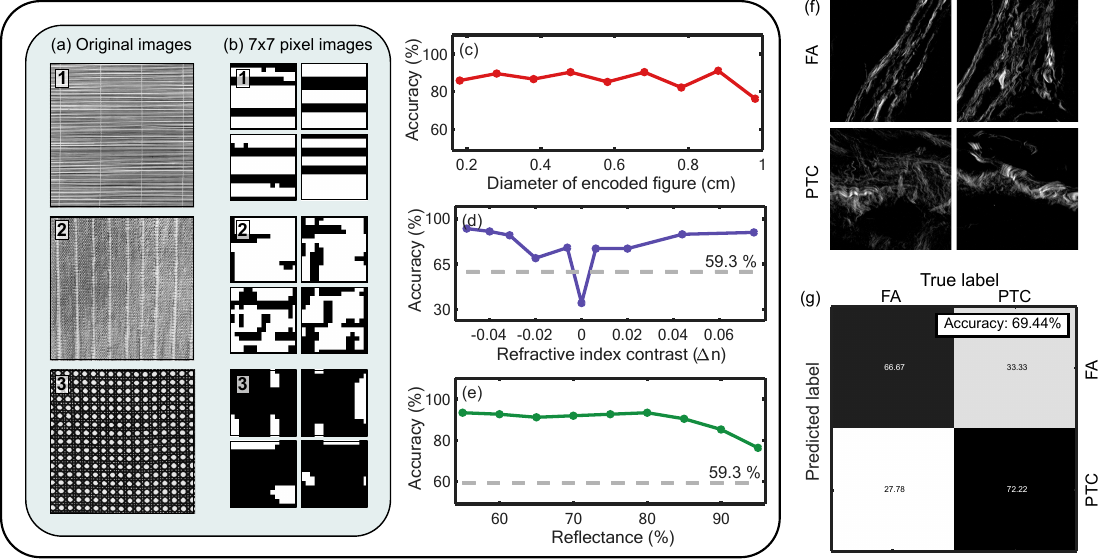}
    \caption{Dependence of tex\-ture-classification performance on the optical system parameters. (a) tex\-ture images with 640\(\times\)640 -pixels for each image. (b) The tex\-ture dataset consisted of three tex\-ture categories, each represented by 150 binary 7\(\times\)7-pixel images. The resulting classification accuracy is plotted as a function of (c) the encoded-image diagonal size, (d) the refractive-index contrast, and (e) the Bragg-mirror reflectivity. The gray dashed line denotes the baseline accuracy obtained using a conventional SVM operating directly on the original 7\(\times\)7-pixel tex\-ture images.(f) Example images from both thyroid collagen categories. (g) Confusion matrix for thyroid collagen classification using the CMPC.}
    \label{fig:classification_accuracy}
\end{figure*}
To evaluate the classification capabilities of the proposed optical reservoir, we first considered a tex\-ture-image data\-set derived from the Brodatz tex\-ture database \cite{Brodatz1966}. Three tex\-ture classes (D49, D85, and D102) were selected, and 150 random 15\(\times\)15-pixel patches were extracted from each 640\(\times\)640-pixel image (see Fig.~\ref{fig:classification_accuracy}(a)), resulting in 450 samples in total.The original 15\(\times\)15-pixel images were resized to 7\(\times\)7 pixels using MATLAB's nearest-neighbor interpolation to match the dimensions used in the previous simulations while preserving the tex\-ture information (see Fig.~\ref{fig:classification_accuracy}(b)). The dataset was divided into training and testing subsets using a 70\%/30\% split. The images were encoded onto the surface of the reservoir in the same manner as described previously. The encoded patterns consisted of pixels, where black pixels corresponded to the refractive index of the reservoir and white pixels corresponded to a perturbed refractive-index value. Speckle patterns generated by the optical system were then classified using a support vector ma\-chine (SVM).

Figure~\ref{fig:classification_accuracy}(c)--(e) presents the classification accuracies obtained under different CMPC parameters. When investigating the influence of the encoded-image size and the Bragg-mirror reflectivity, the refractive-index contrast was fixed at 0.01. For the reflectance analysis, the encoded-image diagonal size was fixed at 0.98 cm. For the refractive-index-contrast analysis, the encoded-image diagonal size and Bragg-mirror reflectance were fixed at 0.98 cm and 95\%, respectively. The gray dashed line indicates the baseline accuracy obtained when the SVM model is directly trained and tested with the original binary images without optical preprocessing. In all cases, the CMPC-based approach outperformed the baseline, indicating that the optical reservoir transforms the input data into a feature space in which the classes are more easily separated by a linear classifier.

Although the tex\-ture-classification task provides a useful proof of concept, the images consisted of only 7$\times$7 pixels and therefore represent a relatively simple classification problem. To assess the applicability of the method to more realistic problems (i.e.~larger images), we next considered a biomedical imaging dataset. Tex\-ture analysis is wide\-ly used in biomedical imaging because tissue tex\-ture and collagen organization often contain diagnostically relevant information \cite{HristuRadu2020Paqo,Hristu2018ThyroidSHG,Bueno2020}. Images were obtained from the PSHG-TISS dataset \cite{Hristu2021PSHGTISS}, which contains polarization-resolved second-harmonic generation (SHG) microscopy images of breast, skin, and thyroid tissues. In this work, only the thyroid collagen subset was used (Fig. \ref{fig:classification_accuracy}(f)). The collagen images exhibit class-dependent variations in several structural and optical properties, including the overall SHG intensity, collagen fiber density, fiber orientation, local intensity fluctuations, and tex\-ture patterns. These characteristics provide diagnostically relevant information and form the basis for distinguishing between the different tissue classes~\cite{Bueno2020}.

The dataset contained two classes: benign follicular adenoma (FA) and malignant papillary thyroid carcinoma (PTC). The original dataset consisted of 40 image sets, with 20 image sets belonging to each class, and each set acquired at ten different polarization angles~\cite{HristuRadu2022PAco}. The final images can be reconstructed from the polarization-resolved images either by averaging all ten polarization images or by using a maximum intensity projection (MIP), in which the highest intensity value across the ten polarization states is selected for each pixel~\cite{HristuRadu2022PAco}. In this study, the images were generated by averaging the measurements obtained at the different polarization angles. To increase the number of samples while maintaining a reasonable computational cost, three random 450\(\times\)450-pixel patches were extracted from each image. 

Prior to simulation, the collagen images were converted into binary representations using the Otsu thresholding method. Pixel values below 15 were treated as background, while the remaining pixels were assigned to the foreground. Small isolated pixels and holes were removed, and the image boundaries were smoothed using a Gaussian filter to reduce high-frequency artifacts. The dataset was then divided into training and testing subsets using the same 70\%/30\% split as in the tex\-ture-image experiments. The baseline classification accuracy achieved using the binary input images was 75\% with SVM. The simulations were calculated with only a single ray. The resulting classification accuracies are shown in Fig.~\ref{fig:classification_accuracy}(g), obtained with a refractive-index contrast of 0.01, a Bragg-mirror reflectivity of 95\%, and an encoded-image diameter of 0.98 cm. Here, with this more complex dataset we were not able to demonstrate optical computing capabilities outperforming the baseline SVM in terms of classification accuracy. We attribute part of the associated challenges to arise from the nature of the considered image data, where important textural information is being stored in greyscale, rather than binarized image data. However, for computational reasons we did not implement greyscale image processing capabilities into our framework. Rather, we were more interested in numerical proof-of-principle demonstration.

Overall, our numerical results demonstrate that the proposed CMPC reservoir can improve classification performance for simple tex\-ture datasets. However, the pres\-ent biomedical imaging task may not fully exploit the capabilities of the CMPC architecture in its current form. The propagation of light inside the reservoir is sensitive to local tex\-ture variations, which is beneficial when the tex\-ture is relatively homogeneous across the encoded image. In the collagen SHG dataset considered here, a significant fraction of the images consists of dark background regions when only the strongest collagen fibers are visible. Consequently, the optical response becomes sensitive not only to the shape of the collagen structures but also to their precise spatial location within the encoded image.

As a result, image patches containing similar collagen fiber patterns may generate noticeably different output intensity fields if the structures are shifted slightly within the image. This sensitivity becomes particularly relevant when data augmentation is used, as cropped image patches can produce substantially different optical responses despite representing the same underlying tissue morphology. Therefore, the classification performance may be limited by the spatially variant nature of the optical transformation rather than by the discriminative capability of the reservoir itself.

In the future, the approach could be extended to grayscale or color images by encoding different pixel values through multiple refractive-index levels, enabling rich\-er optical feature transformations prior to digital classification. Furthermore, the orientation sensitivity of the present architecture could potentially be reduced by illuminating the reservoir simultaneously from multiple directions. In such a configuration, optical interactions would be sampled from all sides of the encoded image, making the resulting output intensity distributions less dependent on image rotation. Consequently, rotationally equivalent tex\-tures would be expected to produce more similar optical responses, reducing the need for rotational data augmentation and potentially improving the robustness of the classification system.

\section{Discussion}

The results demonstrate that an integrated CMPC architecture can perform useful optical feature transformations that increase both the nonlinearity and diversity of image representations. The observed reduction in pairwise cosine similarity suggests that repeated optical interactions distribute information across a higher-dimensional feature space. This interpretation is supported by the classification results: for the texture datasets, the CMPC-generated optical features improved classification performance compared with direct classification of the original binary images, indicating that the reservoir transformation increases class separability.

The parameter study further clarifies which design choices govern the complexity of the optical transformation. Increasing the mirror reflectivity and the encoded-image size generally strengthened the nonlinear behavior and produced more diverse output representations. This is expected, since a higher reflectivity increases the number of round trips each ray undergoes within the reservoir, while a larger encoded image increases the number of refractive-index perturbations the rays interact with. In contrast, the influence of the refractive-index contrast was comparatively weak at the investigated wavelength, although even relatively small index perturbations were sufficient to induce a nonlinear response. This suggests that the presence of perturbations, rather than their magnitude, is the dominant factor in the investigated regime.

At the same time, the simulations reveal several limitations. For example the optical response depends on the spatial position of structures within the encoded image. Consequently, two images containing the same object may generate substantially different output intensity distributions if the object is translated within the field of view.

While image classification provides a convenient benchmark for evaluating the computational properties of the optical reservoir, the observed sensitivity to spatial translations suggests that image recognition may not be its most convenient application domain. For instance, if the position of an object or its edges shifts within the scene, the output signal changes, which is not ideal for classification tasks. In such cases, the system's sensitivity to spatial position could instead be advantageous in simpler tasks aimed at detecting or localizing an object~\cite{RefJ1}. Another possible application would be to utilize the CMPC system for signal processing tasks, such as signal reconstruction or anomaly detection in video, since the reservoir computing paradigm is well suited for temporal prediction tasks~\cite{Cox2026NLPO}.


Several directions remain open for future work. While the present work focuses on binary image encoding, future implementations could incorporate grayscale or color images with multilevel refractive-index modulation. These extensions could increase the information capacity of the reservoir, potentially enhancing performance in complex imaging tasks such as biomedical image analysis and expanding the applicability of the approach to a wider range of mac\-hine-learning problems.

\section{Conclusions}

We have proposed and numerically investigated an integrated photonic CMPC architecture as a platform for optical reservoir computing. Using ray-tracing simulations, we showed that refractive-index perturbations embedded within the reservoir generate complex optical transformations that map input images into high-dimensional optical feature representations.

Characterization using the coefficient of determination ($R^2$) and pairwise cosine similarity showed that the nonlinearity and diversity of the output representations increase with mirror reflectivity and encoded-image size, and that even small refractive-index perturbations are sufficient to induce a nonlinear response. In image-clas\-si\-fi\-ca\-tion tasks on texture datasets, the CMPC-generated features outperformed direct classification of the original binary images.

Overall, these results suggest that integrated photonic CMPC architectures could provide a compact and tunable platform for optical information processing and ma\-chine-learning hardware.

\begin{acknowledgments}
This work was supported by the Jenny and Antti Wihuri Foundation (Grant No. 00250023) and by the Research Council of Finland Flagship Programme, Photonics Research and Innovation (PREIN), Grant No. 346518.
\end{acknowledgments}

\section*{Data Availability Statement}
The data that support the findings of
this study are available from the
corresponding author upon reasonable
request.

\bibliography{Bibliography}

\clearpage
\onecolumngrid
\appendix

\section{Effects of Wavelength on Phase Change}
\label{chap:effects_wavelength}

\setcounter{figure}{0} \setcounter{table}{0} \setcounter{section}{0} \renewcommand{\thefigure}{S\arabic{figure}} \renewcommand{\thetable}{S\arabic{table}} \renewcommand{\thesection}{S\arabic{section}}

The numerical simulations performed during this work systematically showed, that the calculated output intensity fields were highly sensitive to even very small phase variations. At a wavelength of 1550nm, which is commonly used in integrated photonics, the phase rotates very rapidly around the complex plane even when the optical propagation distance is short. In our simulation framework, this resulted in substantial changes in the output signal, which is obtained by rasterizing the rays propagated through the reservoir into a discrete, complex-valued 120\(\times\)120 matrix.

\begin{figure*}[!t] \includegraphics[scale=0.8]{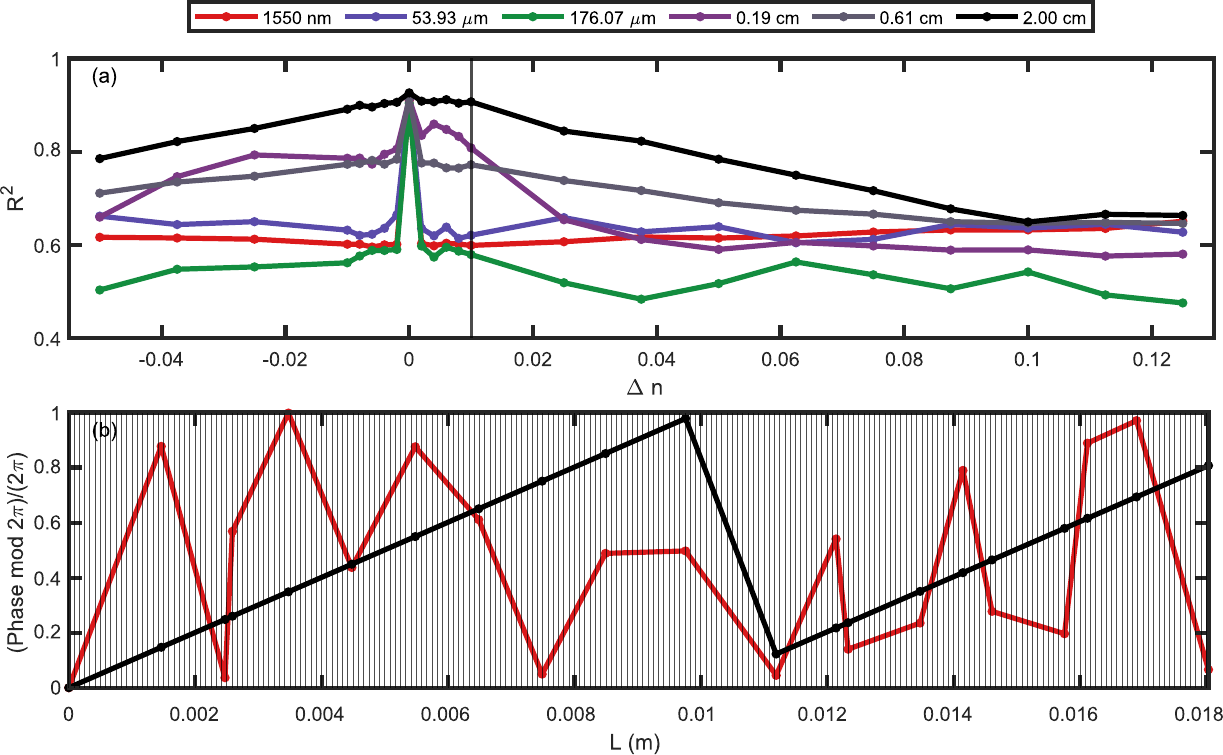} \centering \vspace{-1.5mm} \caption{\label{fig:aallonpituuksien_erot} (a) \(R^2\) curves for different wavelengths as a function of refractive index contrast. (b) Normalized phase for two wavelengths, 1550nm and 2cm, at a refractive index contrast of 0.01. The y-axis represents the phase variation over the range 0 to 2\(\pi\), corresponding to one complete rotation around the complex plane.} \vspace{-3.5mm} 
\end{figure*}

In addition, we repeated the simulations for 130 configurations as a function of refractive index contrast using longer wavelengths while keeping the refractive indices, and hence the ray trajectories, unchanged. As the wavelength increases, the phase approaches a constant value when the optical propagation distance is short. Figure~\ref{fig:aallonpituuksien_erot}(a) shows the \(R^2\) curves for six different wavelengths. We observe that the narrow peak broadens as the wavelength increases. Furthermore, Figure~\ref{fig:aallonpituuksien_erot}(b) presents the phase evolution as a function of propagation distance for two wavelengths in a single simulated configuration with a refractive index contrast of 0.01 (indicated by the vertical line in panel (a)).

In Figure~\ref{fig:aallonpituuksien_erot}, the phase has been normalized to the range between 0 and 1, corresponding to one complete rotation around the complex plane. The vertical lines indicate the width of a single pixel in the output intensity matrix. The effect of wavelength is clearly visible: for shorter wavelengths, the phase varies significantly between adjacent pixels, whereas for longer wavelengths the phase changes more gradually. This result is in line with Fig.~S1a), where the very long (and strictly speaking unrealistic) wavelengths broaden the \(R^2\) peak near $\Delta n = 0$ value. In overall, these simulations show that the peaking of the system's \(R^2\) value near $\Delta n = 0$ is caused by interference effects within the field propagating inside the CMPC. In short, the studied system is, in practise, an interferometric instrument.

 \begin{figure*}[!t] \includegraphics[scale=0.8]{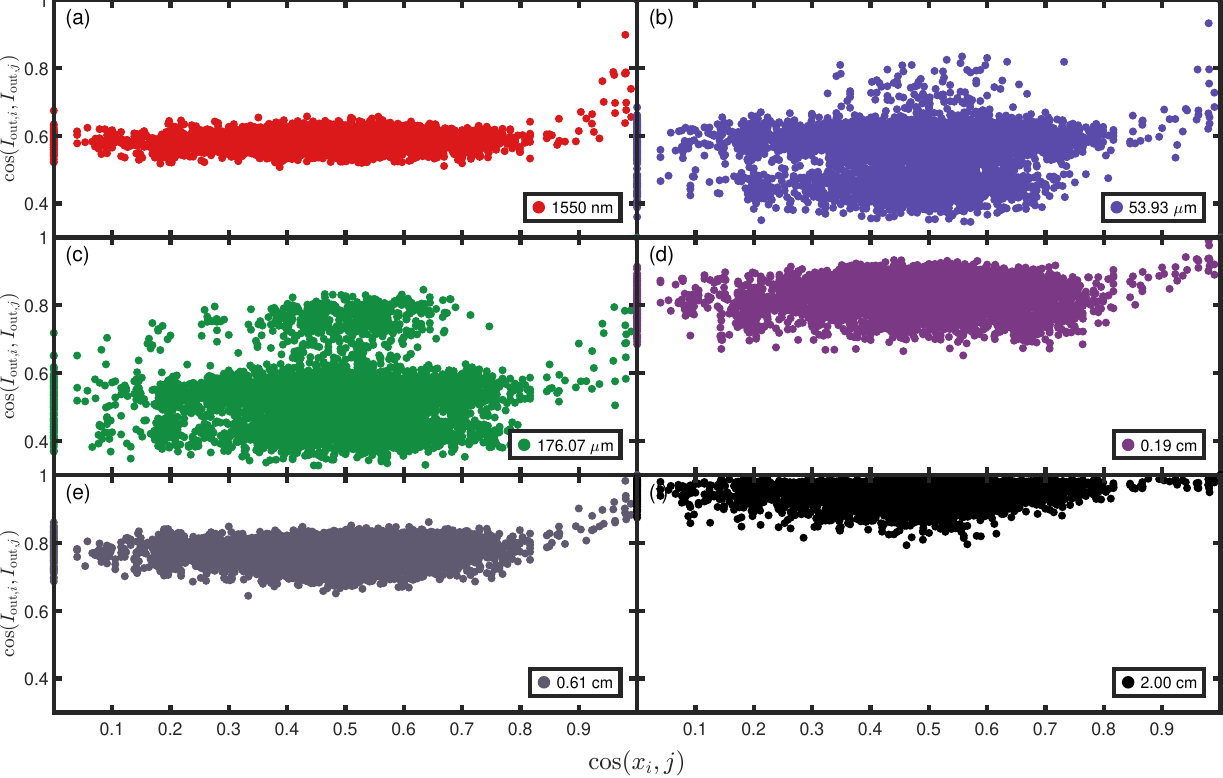} \centering \vspace{-1.5mm} \caption{\label{fig:similarity_aallonpituuksien_erot} Pairwise cosine similarity for the input configurations and output intensity fields at different wavelengths, with an encoded image diameter of 0.98 cm in the simulations, a reflectivity of 95\%, and a refractive index contrast of 0.01. The plots show good agreement with the corresponding \(R^2\) values presented above for different wavelengths: the similarity of the output intensity fields increases as the values of the \(R^2\) curves decrease.} \vspace{-3.5mm} 
\end{figure*} 

Similarly to the article, we examined the pairwise cosine similarity between the simulated output intensity fields at different wavelengths. The fields were calculated for a refractive index contrast of 0.01 (indicated by the vertical line in Figure~\ref{fig:aallonpituuksien_erot}), and the results are presented in Figure~\ref{fig:similarity_aallonpituuksien_erot}. It can be observed that the cosine similarity values follow the trend of the calculated \(R^2\) values reasonably well. Furthermore, Figure~\ref{fig:similarity_aallonpituuksien_erot}(f) suggests that as the phase shift approaches a constant value, the optical path length traveled by the rays may have a stronger influence on the resulting intensity distribution. Consequently, the intensity fields become increasingly similar despite the refractive index contrast. We also note, that we can still achieve a nonlinear response from the system, when the refractive index contrast increases.

\end{document}